\documentclass[a4paper,12pt]{article}
\usepackage{amsmath}
\usepackage{amsthm}
\usepackage{enumitem}
\usepackage{afterpage}
\usepackage[autostyle]{csquotes}
\usepackage{mathtools}
\usepackage{cancel}
\usepackage{latexsym}
\usepackage[latin1]{inputenc}
\usepackage{amsmath}
\usepackage{amsfonts}
\usepackage{adjustbox}
\usepackage{amssymb}
\usepackage{framed}
\usepackage{bbm}
\usepackage{rotating} 
\usepackage{xcolor}
\usepackage{graphicx}
\usepackage{setspace}
\usepackage{dsfont}
\usepackage{bigints}
\usepackage{setspace}
\usepackage{natbib}
\usepackage{booktabs}
\usepackage{multirow}
\usepackage{longtable} 
\usepackage{booktabs}
\usepackage{siunitx}
\usepackage{rotating}

\usepackage{changes}

\newtheorem{prop}{Proposition}
\newtheorem{lem}[prop]{Lemma}
\newtheorem{corollary}{Corollary}

\newtheorem{prop1}{Proposition}[section]
\newtheorem{theorem}[prop1]{Theorem}

\newtheorem{rmk1}{Remark}

\newcommand{\ba}{\begin{eqnarray}}
\newcommand{\ea}{\end{eqnarray}}
\newcommand{\bi}{\begin{itemize}}
\newcommand{\ei}{\end{itemize}}
\newcommand{\ben}{\begin{enumerate}}
\newcommand{\een}{\end{enumerate}}
\newcommand{\blem}{\begin{lem}}
\newcommand{\elem}{\end{lem}}
\newcommand{\bteo}{\begin{theorem}}
\newcommand{\eteo}{\end{theorem}}
\newcommand{\bcor}{\begin{corollary}}
\newcommand{\ecor}{\end{corollary}}

\newcommand{\be}{\begin{equation}}
\newcommand{\ee}{\end{equation}}

\newcommand{\VAR}{\mathrm{VaR}}

\def\Em{\mathbb E}

\def\Cc{\mathcal{C}}
\def\Fc{\mathcal{F}}

\def\Ic{\mathcal{I}}

\begin{document}
\title{\textbf{Deep learning Profit \& Loss}\\
\author{
Pietro Rossi\thanks{%
Prometeia S.p.A. and University of Bologna, Bologna, Italy. Corresponding author: pietro.rossi@prometeia.it} \and
Flavio Cocco\thanks{%
Prometeia S.p.A. and University of Bologna, Bologna, Italy.} \and
Giacomo Bormetti\thanks{%
University of Bologna, Italy} 
}
\date{\today}
}
\maketitle

\begin{abstract}
Building the future profit and loss (P\&L) distribution of a portfolio holding, among other assets, highly non-linear and path-dependent derivatives is a challenging task. We provide a simple machinery where more and more assets could be accounted for in a simple and semi-automatic fashion. We resort to a variation of the Least Square Monte Carlo algorithm where interpolation of the continuation value of the portfolio is done with a feed forward neural network. This approach has several appealing features not all of them will be fully discussed in the paper. Neural networks are extremely flexible regressors. We do not need to worry about the fact that for multi assets payoff, the exercise surface could be non connected. Neither we have to search for smart regressors. The idea is to use, regardless of the complexity of the payoff, only the underlying processes. Neural networks with many outputs can interpolate every single assets in the portfolio generated by a single Monte Carlo simulation. This is an essential feature to account for the P\&L distribution of the whole portfolio when the dependence structure between the different assets is very strong like the case where one has contingent claims written on the same underlying.

\vspace{1cm}
\end{abstract}

\section{Introduction}\label{section:introduction}

The computation of the future profit \& loss distribution of a portfolio is the key step 
to manage risk and to set aside regulatory capital. To perform it, the financial industry 
devised different strategies, ranging from historical full evaluation to parametric modeling, 
linear and quadratic mapping on risk factors, and closed-form analytic 
approximations~\citep{mcneil2015quantitative}. As it is well known, the problem has a 
direct and viable solution if we consider only assets whose future price has a fast analytic solution. 
In this case, we just generate enough trajectories to the desired time horizon and, for each 
generated trajectory we have to compute the present value of the portfolio. To build an 
accurate distribution we need several trajectories of the underlying and, if the portfolio has a 
large number of assets, the computation can still be formidable, but it can be easily parallelised. 
If we are not within this context, and the only way to compute the present value of an asset is 
via Monte Carlo simulation, we are in the unpleasant situation that for each scenario we must 
perform a new Monte Carlo simulation. The situation is even worse if we are dealing with 
strongly path-dependent derivatives -- such as American or Bermuda options -- and highly 
non-linear payoff in multiple dimensions. The problem can easily turn out to be not tractable.

One way out of this deadlock is to resort to techniques inspired by the Least Square 
Monte Carlo (LSM)~\citep{longstaff2001valuing} to estimate, via back propagation, the 
continuation value of the portfolio produced by the optimal strategy~\citep{karlin1975first}. This method,
that usually is performed with polynomial interpolation of the continuation value, has proven 
effective in many situations. The major drawback is that for each asset in the portfolio the 
ideal strategy has to be tailored to the asset and, the polynomial interpolation, when used
with many variables, shows a marked preference for over fitting. 

The approach we follow is along the lines of the LSM method, but with some significant variation. 
Interpolation is done using feed forward neural networks (FFNN)~\citep{bishop1995neural,ferguson2018deeply}. 
The impact on the entire procedure is striking. This cures the curse of dimensionality 
associated with the polynomial interpolators. FFNN flexibility and ability to learn (fit) 
the price of the portfolio is remarkable. Rather than using just one trajectory to propagate 
back the continuation value, from each time horizon we perform a short Monte Carlo with very 
few trajectories and, at each time horizon, we train the network to learn the continuation value.
Once we have the coefficients of the trained network at each $t$ we launch a large simulation 
and compute the desired distribution using our trained networks as proxies of the continuation value. 
The key idea behind, that is the main contribution of this work, is that even with relatively few 
trajectories in the training phase, we can still obtain an unbiased set of FFNN capable to 
produce the correct P\&L distribution. It is worth to remark again that the procedure we 
detail is capable to deal with whole portfolios, rather than just single assets, 
and it seems to be the only feasible approach to the proper reconstruction of the P\&L distribution 
in presence of highly non-linear effects and strong cross-dependence among portfolio components. 
As a final comment, our work contributes to the spurring stream of quantitative research approaching the hedging problem 
by means of machine learning techniques~\citep{cao2019deep,buehler2019deep}.  

\section{The problem at hand}

When dealing with market risk what we are mainly interested in is the P\&L distribution. Given the value $V_t$ for our
portfolio at a future time $t$ the profit or loss is defined as the difference
\begin{equation*}
    D(0, t) V_t - V_0,
\end{equation*}
with $D(0,t)$ the discount factor. This way we compare two quantities originating
at the same time. The quantity $D(0,t)V_t$ is a random variable defined as
\[
    D(0,t)V_t = \Em\left[ D(0, T)V_T\,|\, \Fc_t\,\right] = D(0,t) \Em\left[ D(t, T)V_T\,|\, \Fc_t\,\right],
\]
that is the expected value, conditional to the information known in $t$.

The steps needed to estimate the P\& L distribution are:
\begin{enumerate}
\item sample the event space described by $\Fc_t$
\item for each sampled event, compute $D(0,t) V_t$
\item build the cumulative distribution function (CDF) for $ D(0,t) V_t - V_0$.
\end{enumerate}
Once we take control of the CDF for our portfolio, we can tackle
issues like the computation of risk measures, Value-at-Risk ($\VAR$), Expected Shortfall, expectiles, or the expected
positive/negative exposure, if we are interested in CVA or DVA contributions.

Computation of $V_t$, possibly for a large set of values of $t$ can be a daunting process.
If we do not have available a fast way to perform it, we must resort to Monte Carlo simulation,
but a Monte Carlo nested inside another Monte Carlo is most of the time just too expensive from the computational
point of view. The strategy we pursue in this work is to produce a fast device to compute $V_t$ for every $t$ we are interested in and every possible event in $\Fc_t$.  Then, we will use it to produce the desired CDF. The fast device we are hinting at
is an FFNN taking in input, as regressors, all of the underlyings entering the problem and producing a multivariate fit, one value for each asset making up our portfolio.

\section{An optimal stopping problem}

In a financial world -- to fix ideas we can think about handling a portfolio --
we assume decisions based on what we know. These decisions will have consequences 
according to what will happen in the future but we do not know yet. The main consequence 
is that our decision will generate a cash flow $i$ (positive, negative or null ) and will have some
impact on future cash flows.

To make the reasoning more formal, we look at a time horizon $T$, and break it up into $N$
intervals
\begin{equation*}\label{t-intervals}
    \mbox{now}\; = t_0 < t_1 < \cdots < t_{N-1} < t_N = T = \quad\mbox{portfolio horizon}\,.
\end{equation*}
Let $\boldsymbol{S}_n$ the set of variables describing the situation in $t_n$
and, in loose speech, the filtration $\Fc_n$ can be seen as:
\begin{equation*}\label{n-filtration}
    \Fc_n = \bigcup_{i=0}^n \boldsymbol{S}_i,
\end{equation*}
all of the information available up to $t_n$.  The stochastic nature of the world is described by a transition probability 
 $ p(\,\boldsymbol{S}_n, t_n\,|\,\boldsymbol{S}_{n-1}, t_{n-1}\,)$ to go from the state
 $\boldsymbol{S}_{n-1}$ in $ t_{n-1}$ to the state $\boldsymbol{S}_{n}$ in $ t_{n}$.

A strategy $\phi$ is an adapted function to the filtration. In other words,
it is a decision one takes based solely on past and present information.
The decision we assume in $t_n$ is based on $\Fc_n$, and has
immediate consequences depending on $\boldsymbol{S}_n$ and future (unknown)
consequences. 
The immediate consequence is the generation of a cash flow $i_n$.
Think of an American Put option: based on current knowledge, we decide to exercise and the immediate consequence,
is the payoff that we pocket. The future, unknown consequence is that we forgo the possibility to exercise at a later time, possibly in a more advantageous condition. 

The following discussion, purely a review, relies heavily on material from \citep{karlin1975first,longstaff2001valuing} and is a standard issue in the optimal stopping time literature. Let's call $a_n$ \footnote{In the American Put option example, $a_n$ can take one of either two values $\{h=\text{hold},s=\text{stop}\}$.} the decision we take in $t_n$, $\phi$ the strategy emerging from the
$N$ choices we will make and $D(0, t_n)$ the discount factor to be applied to
a cash flow in $t_n$. The value of the $\phi$ strategy we decide to enact will be:
\begin{equation}\label{strategy-value}
	\Ic( \phi, \boldsymbol{S}_0) = \Em\left[\, \sum_{n=1}^N D(0, t_n) i( \boldsymbol{S}_n, a_n) \right].
\end{equation}

\begin{enumerate}
\item evolution occurs in the interval $[t_{n-1}, t_n)$, and we make our decision 
in $t_n$ having full knowledge of everything that happened up to that point;
\item after we have made our choice, $\boldsymbol{S}_n$ will evolve to $\boldsymbol{S}_{n+1}$ 
with a probability described by the matrix
$ p( \boldsymbol{S}_{n+1}, t_{n+1} \,|\, \boldsymbol{S}_n, t_n)$.
\end{enumerate}

Next step is to build a function that is an upper bound for the value described in 
eq.~(\ref{strategy-value}).  Let $V(\boldsymbol{S}_N)$ the payoff at maturity 
of the contract in exam. It is well known that the optimal solution to this problem
is given by the recursive equation described by:
\begin{eqnarray*}
	f_N(\boldsymbol{S}_N) & = & i(\boldsymbol{S}_N, a_N) \stackrel{def}{=} V(\boldsymbol{S}_N),\quad \forall \boldsymbol{S}_N \nonumber
	\\ f_{n-1}(\boldsymbol{S}_{n-1}) & = & \max_{a_{n-1}} \left( i( \boldsymbol{S}_{n-1}, a_{n-1} ), 
	\Em[\, D(t_{n-1}, t_n) f_n(\boldsymbol{S}_{n})\,|\,\Fc_{n-1}\,]  \right),~\text{for}~n=1,\ldots,N-1\,. \label{bounds}
\end{eqnarray*}
The previous solution was first derived in~\citep{bellman1952theory} as the optimality condition in dynamic programming and it is experiencing a new life as the key formula in Reinforcement Learning~\citep{cao2019deep}. 

\section{Working on a portfolio}
Different assets, say $K$, in a portfolio are characterized by the fact that they have different cash flow structure
$i^k_n$  and the previous result can be easily generalized as:
\begin{eqnarray}
	f^k_N(\boldsymbol{S}_N) & = & i^k(\boldsymbol{S}_N, a^k_N) \stackrel{def}{=} V^k(\boldsymbol{S}_N),\quad \forall \boldsymbol{S}_N \nonumber
	\\ f^k_{n-1}(\boldsymbol{S}_{n-1}) & = & \max_{a^k_{n-1}} \left( i^k( \boldsymbol{S}_{n-1}, a^k_{n-1} ), 
    \Em[\, D(t_{n-1}, t_n) f^k_n( \boldsymbol{S}_{n}) | \Fc_{n-1}\,] \right),~\text{for}~n=1,\ldots,N-1\,. \nonumber
\end{eqnarray}
From the above section we conclude that what we need is a reliable estimate of the transition probability
\begin{equation}
 p( \boldsymbol{S}_{n+1}, t_{n+1} \,|\, \boldsymbol{S}_n, t_n). \nonumber
\end{equation}
Even though modeling the right process is the major concern when pricing a portfolio, in this paper we are
more focused on the methodological aspects.  We will make our life simpler and assume that all of the assets
undergo a log-normal process. More complex dynamics can be readily dealt with, provided they belong to the class of Markov processes, possibly  of order higher than one. 

Let $\boldsymbol{S}$ denote the vector of underlying prices $S_1, \ldots, S_K$. Starting from 
$t_{N-1}$, for each trajectory $j$  we have a value $\boldsymbol{S}(j,t_{N-1})$.    
 From each $\boldsymbol{S}(j,t_{N-1})$ we launch $M$ one-step trajectories from $t_{N-1}$ to $t_{N}$  according to
 the law $p(\boldsymbol{S}_{N}, t_{N}\,|\,\boldsymbol{S}_{N-1}, t_{N-1})$. The small set of trajectories originating from
$\boldsymbol{S}(j, t_{N-1})$ will be denoted as $\boldsymbol{S}^m(j, t_{N})$ for $m=1,\ldots,M$.
 In $t_{N}$, we know what is the payoff for each possible triplet $\boldsymbol{S}^m(j, t_N)$ so we can easily compute
 the corresponding payoffs $V^k( \boldsymbol{S}^m( j, t_N)\,)$ for each asset $k=1,\ldots,K$.
The price in $t_{N-1}$ conditioned on $\boldsymbol{S}(j,t_{N-1})$ is given by
\begin{equation}
    V^k( t_{N-1}\,|\,\boldsymbol{S}(j, t_{N-1}))  = \Em[\, D(t_{N-1}, t_N)\,V^k( \boldsymbol{S}( j, t_N)\,)\,|\, \boldsymbol{S}(j, t_{N-1})\,] \label{Vk-def}
\end{equation}
and the right hand side of the above equation can be estimated by the quantity
\begin{equation}
    \frac{1}{M}\sum_{m=1}^M D( t_{N-1}, t_N) V^k( \boldsymbol{S}^m( j, t_N)\,) \label{m-payoff}.
\end{equation}
Now we have $V^k( t_{N-1}\,|\,\boldsymbol{S}(j, t_{N-1}))$ for $k=1,\ldots,K$ at each point $\boldsymbol{S}(j, t_{N-1})$ and we use a neural network to  fit it using as input variables $\boldsymbol{S}( j, t_{N-1})$. We name the interpolating function $\Cc_{int}^{N-1}$ . The symbol $\Cc$ has been chosen as a mnemonic  for ``Continuation'' and is basically a vector value function:
\begin{eqnarray*}
 \Cc^{N-1}_{int} : \boldsymbol{S}(j, t_{N-1}) \rightarrow \boldsymbol{V}( t_{N-1} | \boldsymbol{S}(j,t_{N-1}) ),
\end{eqnarray*}
where $\boldsymbol{V}(t_{N-1}\,|\,\boldsymbol{S}(j, t_{N-1}\,)$ is the vector of prices subject to the fact that we have not exercised our option
till $t_{N-1}$. Components of $\boldsymbol{V}$, defined in eq.~(\ref{Vk-def}), are the price processes of every single asset making up the portfolio.
At this point we repeat the process at $t_{N-2}$. We launch again a small set of one-step trajectories from each
$\boldsymbol{S}(j, t_{N-2}) $ producing $\boldsymbol{S}^m(j, t_{N-1})$.
The difference this time is that in $t_{N-1}$ instead of using the known payoff to compute prices for the
one-step trajectory, we will use the interpolating function $\Cc^{N-1}_{int}$ and compute the price of the trajectory as 
\[ \max( i^k(\boldsymbol{S}^m(j, t_{N-1})), \Cc^{N-1}_{int}(\boldsymbol{S}^m(j, t_{N-1}))\,)  \]
where $i^k$ is the cash flow obtained by exercising the option. The recursive structure of this scheme is rather evident and we can build interpolating functions all the way down to the first exercise date $t_1$.
At each time slice, we have calibrated an interpolating function that we can use to compute continuation values. 
Now, generating P\&L distributions is a straightforward matter:
\begin{enumerate}
\item generate $L$ trajectories from $t_0$ to $t_n$;
\item for each trajectory use the interpolator $\Cc^{n}_{int} $ to compute $V^j(t_n)$;
\item the sample distribution of $D(0, t_n)V^j(t_n) - V(t_0)$ will provide the wanted CDF of the portfolio P\&L.
\end{enumerate}
As a non negligible bonus, we obtain all the marginal distributions for each asset in the portfolio.

The continuation value, propagated backward, is what we use to build the P\&L distribution
and it could be used as a proxy for the price itself. Unfortunately, it is not a reliable neither an accurate quantity. 
For once, the numbers obtained
can hardly be considered independent. Therefore we cannot produce at all an estimate for the statistical error.
Besides, as is well known in literature~\citep{longstaff2001valuing} such a procedure produces a value biased towards higher 
values. For accurate pricing, it is a much better procedure to use our interpolator as our policy maker. 
We generate a brand new set of trajectories and for each trajectory we use our FFNN-based continuation 
value as the strategy maker. For each time horizon we decide to exercise if the payoff is higher than the 
continuation value, otherwise we check the next time slice. It is worth pointing out that this way of proceeding
establishes a strategy that is not looking forward, therefore it can be, at best, as good as the optimal. 
The price computed this way is always a lower bound for the real price. It is a simple consequence of the fact that 
any legitimate strategy is at most as good as the optimal strategy.

\section{Numerical Experiments}
In all the numerical experiments performed, the stochastic process of the underlying
is defined by:
\begin{equation}
    S_i(t+\Delta t) = S_i(t) \exp\left( ( r - \delta_i)\Delta t - \frac{\sigma^2_i}{2}\Delta t + \sigma_i (\; W_i(t + \Delta t) - W_i(t)\;) \right), \nonumber
\end{equation}
with $r $ the instantaneously short rate, $\delta_i$ the continuously compounded dividend yield, $\sigma_i$ the volatility of the
Wiener process $W^i_t$.  All Wiener processes are non correlated. Furthermore
in modeling the portfolios used as an example, we have made some technically simplified assumption, namely
all of the assets within a given portfolio share the same exercise schedule and maturity.

\subsection{Portfolio Composition}

The portfolio we study is made up of three assets:
\begin{enumerate}
\item an American Put written on $S_x$, whose payoff reads
\[
    \text{payoff}_{AM} = (\, \kappa_{am} - S_x(T) \,)^+ ,
\]
\item a European Call on Min written on $S_x$ and $S_y$, where
\[
    \text{payoff}_{Cm} = (\, \min( S_x(T), S_y(T) ) - \kappa_{cm} \,)^+,
\]

\item a Bermuda Call on Max written on $S_x$, $S_y$, and $S_z$, whose payoff is given by
\[
    \text{payoff}_{bCM} = (\, \max( S_x(T), S_y(T), S_z(T) ) - \kappa_{bcm} \,)^+.
\]
\end{enumerate}

In section \ref{one-year} we look at a one year maturity portfolio with monthly exercise schedule
both for the American Put and the Bermuda option, while in section \ref{three-year} we look at the same portfolio 
on a three years horizon and four months exercise schedule.

\subsection{The Interpolator}
\begin{figure}[ht] \centering 
\includegraphics[scale=0.3]{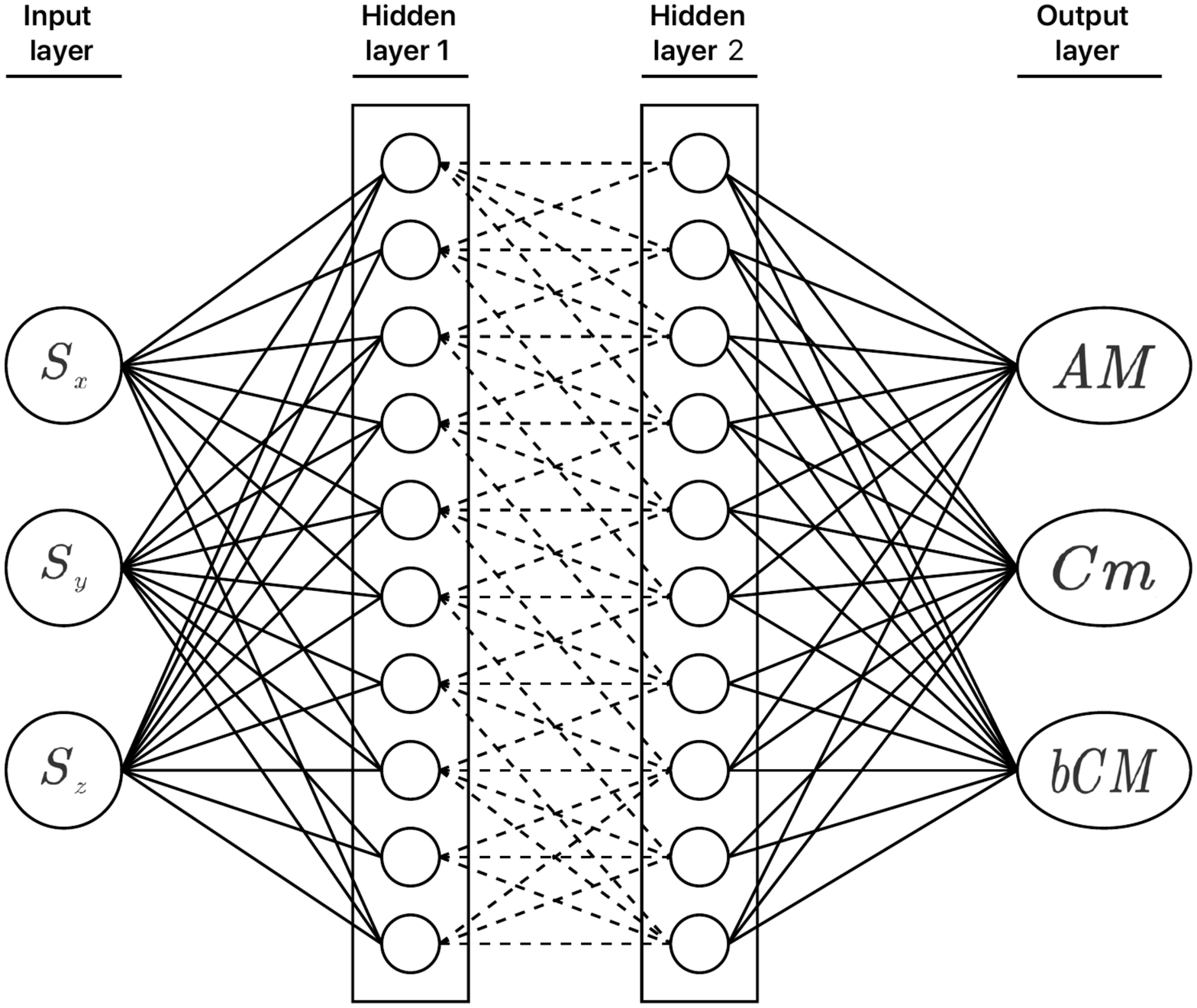}
  \caption{\label{fig:nnet}
   The topology of the FFNN used as interpolator at each exercise date. All the nodes in the hidden layers are connected even though,
      for graphical reason, this is not properly represented. For the same reason, the bias node at the input layer and each of the
      hidden layers, is not represented. The output nodes will produce, as described in the text, the continuation
      value of each of the assets in the portfolio. The legenda, from top down is as follow: American Put, European Call on Min, Bermuda Call on Max.
  } \end{figure}

The network used for the interpolation is a very simple feed forward network with 
two hidden layers, with 10 nodes each. Its topology is detailed in fig.(\ref{fig:nnet}).  The activation functions are sigmoidals on both layers.
We have tried with different activation functions without any visible benefit.

\subsection{Experiments and checks}
The portfolio described above has been simulated in two different contexts: the first one for a one-year maturity and
monthly exercise schedule, the second for a three-year maturity and triannual exercise schedule.
In sections (\ref{one-year}, \ref{three-year}) we show results for the price of the portfolio in these two scenarios and the
quantiles for each individual assets as well as the whole portfolio. Given our definition of P\&L negative quantiles correspond to losses,
then VaR is the negative of the quantile. To check for correctness of the results is not very simple. 
We elected to perform the following checks. As far as the individual assets are concerned, we compared results,
both for prices  and quantiles, obtained while handling the whole portfolio in one simulation, with
results obtained simulating each asset individually. Results show no difference in the two cases.

\begin{figure}[ht] \centering 
  \includegraphics[scale=.35]{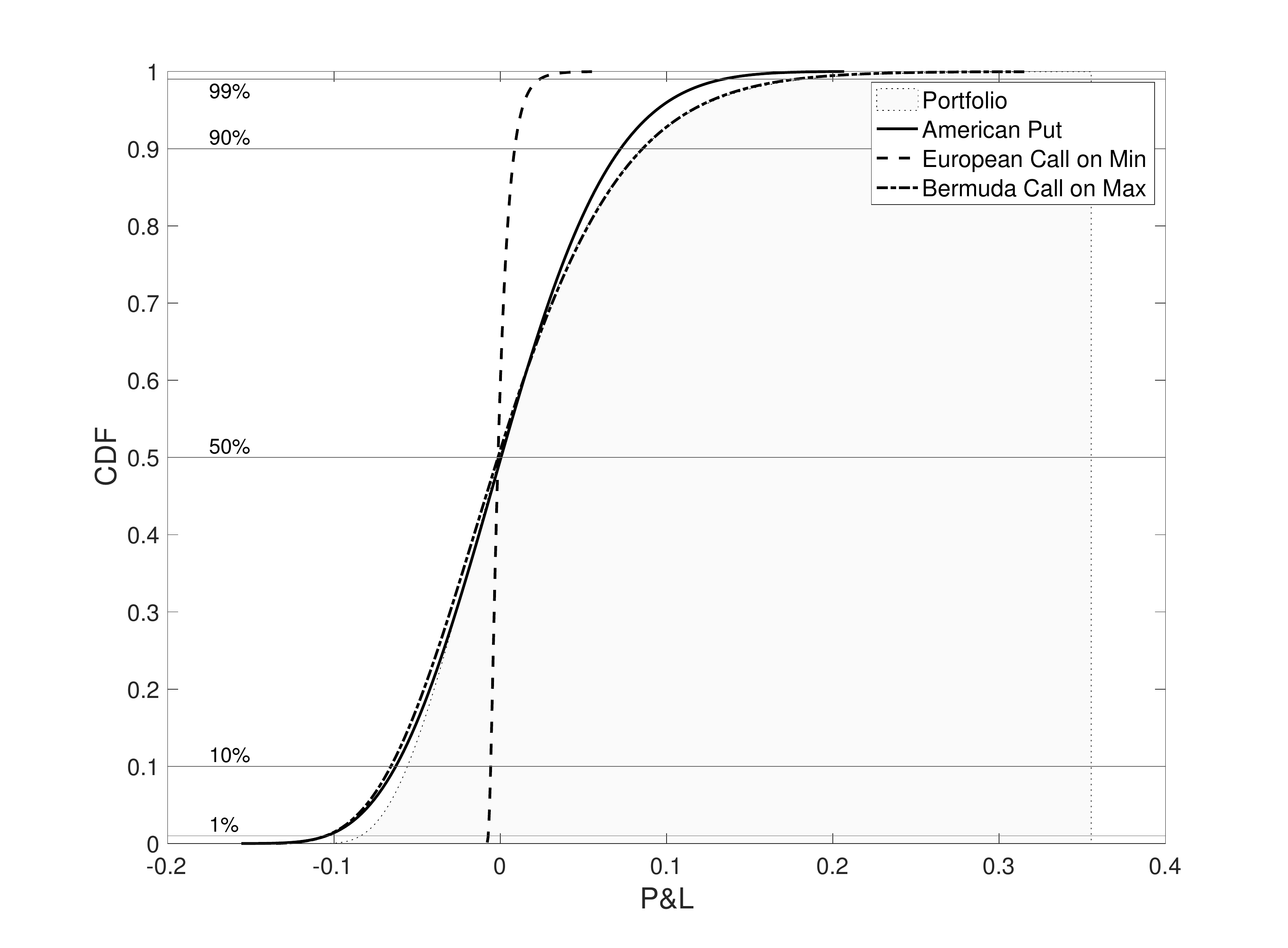} 
  \caption{\label{fig:fourmonths}
    Three years portfolio, four months after contract start. The filled gray curve corresponds to the P\&L CDF of the whole portfolio. The bold, dashed, and dashed-dotted lines represent the P\&L CDF of the American Put, European Call on Min, and Bermudan Call on Max, respectively. The CDF of each single asset was computed while processing the whole portfolio.
  } \end{figure}

\begin{figure}[ht] \centering 
  \includegraphics[scale=.35]{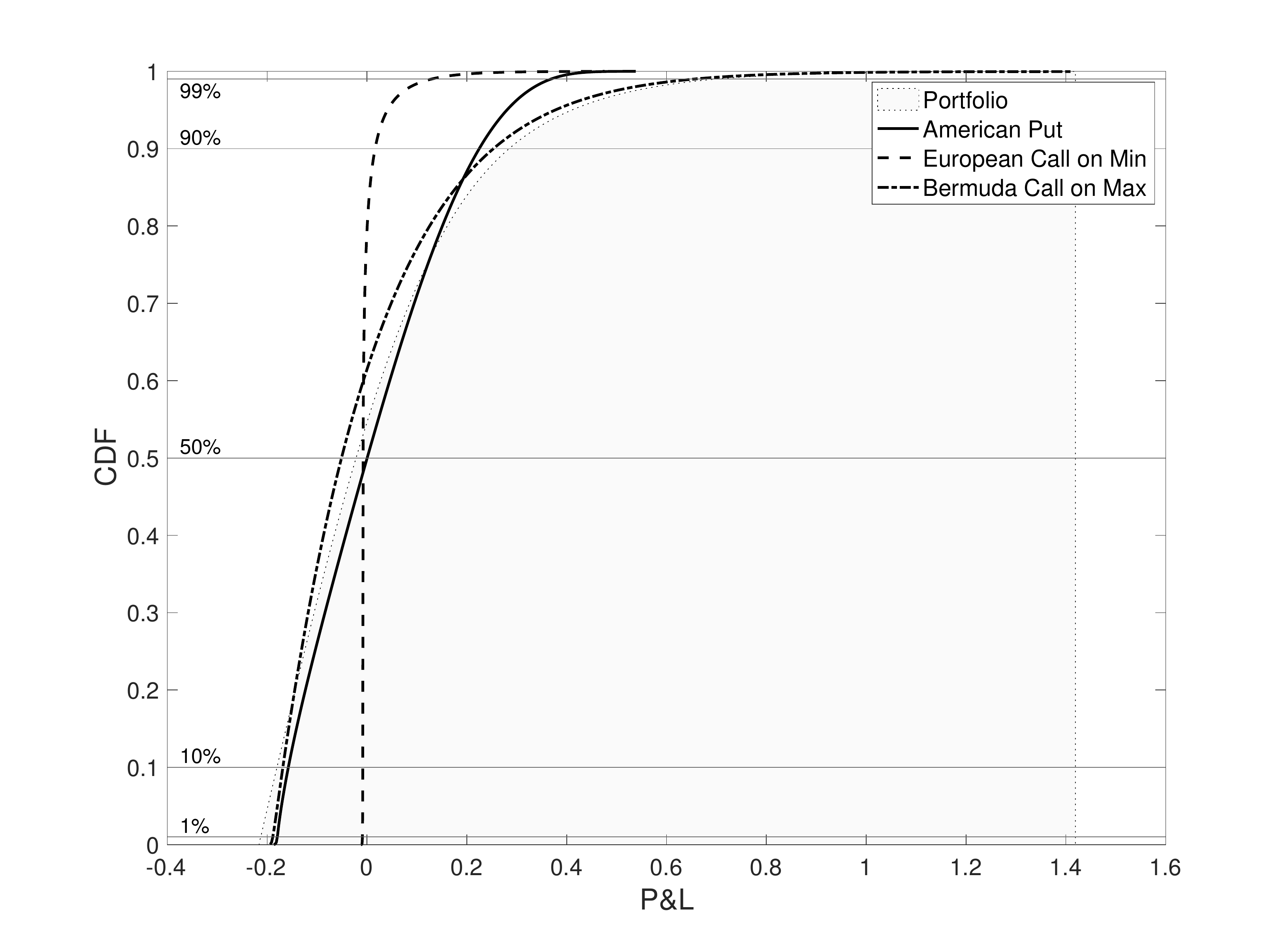} 
  \caption{\label{fig:twoyears}
        Three years portfolio, two years  after contract start. Legend as in caption of figure~(\ref{fig:fourmonths}). 
            \label{fig_ii}
  } \end{figure}

For some of the assets considered there are in the literature benchmark results, namely the Bermuda Call on Max 
and the American Put can be compared with high precision values coming from PDE approaches. Results form both
of these checks are presented in section~\ref{check}.
Figures~(\ref{fig:fourmonths}) and~(\ref{fig:twoyears}) come from the three-year portfolio. They compare the P\&L distribution 
four months after contract start and two years after contract start. The comparison is done for each single asset individually as well as the whole portfolio. The filled gray curve corresponds to the P\&L CDF of the whole portfolio. The bold, dashed, and dashed-dotted lines represent the P\&L CDF of the American Put, European Call on Min, and Bermudan Call on Max, respectively. It is important to stress that the CDF of each asset was computed concurrently with the P\&L distribution of the entire portfolio. For ease of readability, both figures report the horizontal lines corresponding to the 1, 10, 50, 90, and 99 percent probability levels. The Tables~(\ref{tab:fourmonths}) and~(\ref{tab:twoyears}) detail the associated quantile values for the portfolio and for each single component.  

The procedure is capable to provide accurate results and the ability to deal with whole portfolios, rather than just single assets
allows for a rather accurate estimate of the hedging effects. In this toy example  we have clearly the American Put hedging against the two calls, 
and the results of the VaR show markedly this effect. It is worth pointing out that the VaR of the whole portfolio could not have been 
estimated by any approach that would consider each asset separately. The portfolio was build with assets hedging each other and the only way to account for this effect is to compute, for each scenario, the contribution to the P\&L distribution of the whole portfolio.
This hedging effect is quite noticeable both in table \ref{q-1m} than in table \ref{q-1Y}. The portfolio $\VAR$ is significantly different than the sum of $\VAR$ for each asset.

\subsection{One-year maturity}\label{one-year}
The maturity is one year after contract start, exercise dates are monthly, both for the American Put and the
Bermuda Call on Max. Parameter values are as follows

\begin{equation}
    S_x = 1.0, \quad S_y = 1.0, \quad S_x = 1.0, \nonumber
\end{equation}

\begin{equation}
    \kappa_{am} = 1.0, \quad \kappa_{cm} = 0.9, \quad \kappa_{bCM} = 1.0, \nonumber
\end{equation}

\begin{equation}
    \sigma_x = \sigma_y = \sigma_z = 0.2, \quad r = 5.0\%, \quad \delta_x = \delta_y = \delta_z = 3.0\%\,. \nonumber
\end{equation}

\subsubsection{Summary of Results}  

  \begin{table}[h!] \centering
  \begin{tabular}{l|r} 
                    Asset & Price
                \\ \hline
                   AM        &   6.943  $ \pm $  0.003
                \\ Cm        &   5.876  $ \pm $  0.003
                \\ bCM       &  19.518  $ \pm $  0.005
   \end{tabular}
    \caption{ 
            The price of each asset within the portfolio. The maturity of every asset
            is one year and the exercise schedule is monthly. 
           }
   \end{table}


  \begin{table}[h!] \centering
  \begin{tabular}{l|r|r|r|r|r} 
                    Asset/Quantile & .01 & .10 & .50 & .90 & .99 
                \\ \hline
                   portfolio &  -8.08 & -4.79 & 0.07  & 5.12  & 8.80
                \\  \hline
                   AM        &  -4.01 & -2.77 &  .27  & 3.49  & 7.17
                \\ Cm        &  -3.63 &  2.37 &  .15  & 2.47  & 3.78
                \\ bCM       &  -7.39 & -4.26 &  .11  & 4.40  & 7.24
   \end{tabular}
  \caption{ 
             Quantiles of the P\&L distribution one months after contract start
             for a portfolio with one-year maturity and monthly exercise schedule.
             Quantiles are given for the whole portfolio as well as each single asset.
    \label{q-1m}
           }
   \end{table}


  \begin{table}[h!] \centering
  \begin{tabular}{l|r|r|r|r|r} 
                    Asset/Quantile & .01 & .10 & .50 & .90 & .99 
                \\ \hline
                   portfolio & -15.36 & -10.18 & -.50   & 13.51 &  26.34
                \\  \hline
                   AM        &  -5.92 & -5.21 & -1.00 &  7.40 & 16.25
                \\ Cm        &  -5.47 & -4.35 &  -.86 &  5.94 & 12.75
                \\ bCM       & -13.81 & -8.92 &  -.29 & 10.93 & 20.30
   \end{tabular}
  \caption{ 
             Quantiles of the P\&L distribution six months after contract start
             for a portfolio with one-year maturity and monthly exercise schedule.
             Quantiles are given for the whole portfolio as well as each single asset.
    \label{q-1Y}
           }
   \end{table}

\subsection{Three-year maturity}\label{three-year}

The maturity is three years after contract start, exercise dates are every four months, both for the American Put and the
Bermuda Call on Max. Parameter values are as follows

\begin{equation}
    S_x = 1.0, \quad S_y = 1.0, \quad S_z = 1.0, \nonumber
\end{equation}

\begin{equation}
    \kappa_{am} = 1.0, \quad \kappa_{cm} = 1.0, \quad \kappa_{bCM} = 1.0, \nonumber
\end{equation}

\begin{equation}
    \sigma_x = \sigma_y = \sigma_z = 0.2, \quad r = 5.0\%, \quad \delta_x = \delta_y = \delta_z = 10.0\%\,. \nonumber
\end{equation}

\subsubsection{Summary of Results}  

  \begin{table}[h!] \centering
  \begin{tabular}{l|r} 
                    Asset & Price
                \\ \hline
                   AM        &  18.020  $ \pm $  0.005
                \\ Cm        &   0.844  $ \pm $  0.002
                \\ bCM       &  18.666  $ \pm $  0.006
   \end{tabular}
    \caption{ 
            The price of each asset within the portfolio. 
            The maturity of every asset 
            is three years and triannual exercise schedule.  
            All assets have been computed concurrently 
            using an interpolating network with three output units.
           }
   \end{table}

  \begin{table}[h!] \centering
  \begin{tabular}{l|r|r|r|r|r} 
                    Asset/Quantile & .01 & .10 & .50 & .90 & .99 
                \\ \hline
                   portfolio &  -8.32  & -5.55 & -0.03 &  8.61 & 18.35
                \\  \hline
                   AM        & -10.37 & -6.26 &  0.07 &  7.22 & 13.43
                \\ Cm        &  -0.75 & -0.58 & -0.13 &  0.87 &  2.38
                \\ bCM       & -10.56 & -6.56 & -0.12 &  8.54 & 17.61
   \end{tabular}
  \caption{  \label{tab:fourmonths}
             Quantiles of the P\&L distribution four months after contract start
             for a portfolio with three-year maturity and triannual exercise schedule.
             Quantiles are given for the whole portfolio as well as for each asset.
             All assets have been computed concurrently 
             using an interpolating network with three output units.
           }
   \end{table}

  \begin{table}[h!] \centering
  \begin{tabular}{l|r|r|r|r|r} 
                    Asset/Quantile   &  .01     & .10     & .50     & .90    & .99 
               \\ \hline
                  portfolio          &  -21.23  &  -17.93 &  -2.14  &  28.74 & 69.53 
               \\  \hline
                   AM                &  -17.94  &  -15.64 &   0.01  &  22.41 & 36.83
               \\  Cm                &  -0.93   &  -0.85  &  -0.76  &  1.58  & 12.92
               \\  bCM               &  -18.49  &  -16.81 &  -5.13  &  25.51 & 64.65

   \end{tabular}
   \caption{ \label{tab:twoyears}
             Quantiles of the P\&L distribution two years after contract start
             for a portfolio with three-year maturity and triannual exercise schedule.
             Quantiles are given for the whole portfolio as well as for each asset.
             All assets have been computed concurrently 
             using an interpolating network with three output units.
           }
   \end{table}

\clearpage

\section{Comparison with existing results}\label{check}
 There are no results listed in the literature for the quantiles of the future P\&L distribution but,
 for some of the parameters used above, namely for the three-year exercise schedule,
 some results are provided in~\citep{becker2019deep,goudenege2020machine},
 concerning the Bermuda Call on Max. 
 
 Table~(\ref{bCM-cv}) compares our results with those in~\citep{becker2019deep} where the price and the $95\%$ confidence interval (CI) are provided. These results have been obtained using the interpolated continuation value as a proxy for the exercise surface. FFNN-based prices are fully consistent with the benchmark 95\% CI. 

   \begin{table}[h!] \centering
   \begin{tabular}{l|r|r|r|c} 
                     Nr Asset   & $S_0$ & price $\pm$ 95\% MC error  & results from~\citep{becker2019deep} &  95\% CI
                \\  \hline
                    2   &  90 & $  8.071 \pm  0.005 $ &    8.074 & [ 8.060,8.081] 
                \\  2   & 100 & $ 13.901 \pm  0.007 $ &   13.899 & [13.880,13.910] 
                \\  2   & 110 & $ 21.345 \pm  0.006 $ &   21.349 & [21.336,21.354] 
                \\  3   &  90 & $ 11.275 \pm  0.007 $ &   11.287 & [11.276,11.290] 
                \\  3   & 100 & $ 18.683 \pm  0.008 $ &   18.690 & [18.673,18.699] 
                \\  3   & 110 & $ 27.562 \pm  0.007 $ &   27.573 & [27.545,27.591] 
    \end{tabular}
    \caption{\label{bCM-cv} Bermuda Call on Max prices and Monte Carlo errors from FFNN compared with results from~\citep{becker2019deep}.}
    \end{table} 
 
Furthermore, we can perform accurate checks for the  American Put option. In table~(\ref{lsmvspde}) we give the corresponding results.  
The column labeled $\Delta t$ is the interval between exercise dates, and the value $\Delta t = 0$
is the result of a linear regression on the non zero $\Delta t$. The column labeled PDE
provides the result coming from a method based on partial differential equations.  The PDE result
has been obtained solving the equation with a semi-implicit method, both with an SOR and an FFT 
based preconditioning. Some results with dividends are shown in table~(\ref{lsmvspde+div}). The quality of the agreement
between the FFNN-based values and the PDE results is remarkable. 

   \begin{table}[h!] \centering
   \begin{tabular}{l|rrrl|r} 
                     $\Delta t$ & $S_0$ & Price &       & Err   & PDE
                \\  \hline
                    2M          &  90 & 11.340 & $\pm$ & 0.003
                \\  1M          &  90 & 11.416 & $\pm$ & 0.003
                \\  2W          &  90 & 11.454 & $\pm$ & 0.003
                \\  1W          &  90 & 11.472 & $\pm$ & 0.003
                \\   0          &  90 & 11.489 &       &       & 11.490
                \\  \hline
                    2M          & 100 &  5.997 & $\pm$ & 0.003 &
                \\  1M          & 100 &  6.041 & $\pm$ & 0.002 &
                \\  2W          & 100 &  6.066 & $\pm$ & 0.003 &
                \\  1W          & 100 &  6.075 & $\pm$ & 0.003 &
                \\   0          & 100 &  6.086 &       &       & 6.089
                \\  \hline
                    2M          & 110 &  2.936 & $\pm$ & 0.003 &
                \\  1M          & 110 &  2.958 & $\pm$ & 0.002 &
                \\  2W          & 110 &  2.972 & $\pm$ & 0.003 &
                \\  1W          & 110 &  2.978 & $\pm$ & 0.003 &
                \\  0           & 110 &  2.983 &       &       & 2.985
                \\  \hline
    \end{tabular}
    \caption{ r  = 5\%, $\sigma$ = 20\%,  T = 1 year, Strike= 100, $\delta = 0 $. \label{lsmvspde}}
    \end{table}

   \begin{table}[h!] \centering
   \begin{tabular}{l|rrrl|r} 
                     $\Delta t$ & $S_0$ & Price &       & Err   & PDE
                \\  \hline
                    2M          &  90 & 12.309 & $\pm$ & 0.003
                \\  1M          &  90 & 12.348 & $\pm$ & 0.006
                \\  2W          &  90 & 12.370 & $\pm$ & 0.004
                \\  1W          &  90 & 12.377 & $\pm$ & 0.004
                \\   0          &  90 & 12.387 &       &       & 12.384
                \\  \hline
    \end{tabular}
    \caption{ r  = 5\%, $\sigma$ = 20\%,  T = 1 year, Strike= 100, $\delta = 3\% $. \label{lsmvspde+div}}
    \end{table}


\section{Conclusion}\label{sec:conclusion}

We presented a general and simple machinery to compute the future P\&L distribution of a portfolio including non-linear, path-dependent, and strongly correlated derivatives. The idea was to leverage the flexibility of feed forward neural networks as universal approximators and to employ them in a Least-Square Monte Carlo approach to price American and Bermuda derivatives. The advantages of the approach are manifold: i) the machinery can easily manage the inclusion of new instruments. ii) The FFNNs cure the drawbacks related with the curse of dimensionality, which are inherently a problem with polynomial approximating functions. iii) The neural networks are better designed to deal with non differentiable payoffs. iv) It is the unique viable approach for concurrently pricing instruments with different exercise style and sensitive to the same risk factors avoiding the computationally burden nested Monte Carlo procedure. In this respect, it is worth stressing once more that our approach jointly recovers the portfolio P\&L distribution and the single instruments' marginals.

This paper details the results from a toy experiment where we considered a portfolio composed by three strongly dependent derivatives -- an American Put, a European Call on Min, and a Bermuda Call on Max -- whose underlying assets follow a simple dynamics. As a future perspective, we plan to investigate more realistic dynamics in a higher dimensional setting. Specifically, we want to assess to which extent the approach can be extended to manage non-Markov dynamics. This case is of particular relevance given the flourishing of stochastic models where the volatility is driven by a fractional dynamics. The interplay among the long-memory features and the high-dimensional nature of a portfolio may result in a mixture whose complexity can be cured only by resorting to a neural network approach.

\end{document}